\documentstyle[12pt]{article}
\textheight by \topskip \textwidth 450pt

\newcommand{\be}{\begin{equation}}\newcommand{\ee}{\end{equation}}
\newcommand{\bea}{\begin{eqnarray}}\newcommand{\eea}{\end{eqnarray}}
\newcommand{\brr}{\begin{array}}\newcommand{\err}{\end{array}}
\newcommand{\ben}{\begin{enumerate}}\newcommand{\een}{\end{enumerate}}

\newcommand{\ba}{\begin{array}}
\newcommand{\ea}{\end{array}}

\def\lab{\label}
\def\lf{\left}

\def\non{\nonumber}\def\pa{\partial}\def\ran{\rangle}
\def\rar{\rightarrow}\def\Rar{\Rightarrow}
\def\ri{\right}\def\ti{\tilde}
\def\al{\alpha}\def\bt{\beta}
\def\de{\delta}\def\De{\Delta}
\def\te{\theta}
\def\si{\sigma}
\def\om{\omega}
\newcommand{\mlab}[1]{\label{#1}}

\def\1{{_{1}}}\def\2{{_{2}}}
\begin{document}

\begin{center}

{\large \bf Group theoretical aspects of neutrino mixing \\[3mm]
in Quantum Field Theory}

\vspace{1cm}

M.~Blasone$^{1,2}$, A.~Capolupo$^{1}$ and G.~Vitiello$^{1}$

\vspace{.4cm}

$^1$ Dipartimento di Fisica and INFN, Universit\`a di Salerno, 84100
Salerno, Italy

$^2$ Institute f\"ur Theoretische Physik, Freie Universit\"at Berlin,
 \\ Arnimallee 14, D-14195 Berlin, Germany
\end{center}

\vspace{.5cm}

\small

\centerline{\bf Abstract}

By resorting to recent results on the Quantum Field Theory of mixed
particles, we discuss some aspects of three flavor neutrino mixing.
Particular emphasis is given to the related algebraic structures and their
deformation in the presence of CP violation.
A novel geometric phase related to CP violation is introduced.

\normalsize
\section{Introduction}

Some progress has been done recently in the direction of finding
a proper mathematical setting for
the description of mixing in Quantum Field Theory (QFT).
This is obviously a quite relevant task, in view of the importance of
neutrino and meson mixing in the context of particle physics
\cite{Cheng-Li,experiments}.

It is worth to point out that \cite{Barg}  mixing of states with
different masses is not even  allowed
in non-relativistic Quantum Mechanics (QM). In
spite of this fact, the quantum mechanical treatment \cite{Pont}
is the one
usually adopted for its simplicity and elegance. A review of the
problems connected with the QM treatment of mixing and
oscillations can be found in Ref.\cite{Zralek}. Difficulties in
the construction of the Hilbert space for mixed neutrinos were
pointed out in Ref.\cite{kimbook}.

Only recently \cite{BV95,3flavors,binger,fujii,hannabuss} a consistent treatment of
mixing and oscillations in QFT has been achieved, based on the discovery
of a  rich non-perturbative  structure  associated to the
vacuum for mixed particles. This vacuum appears to be a condensate of
particle-antiparticle pairs, both for fermions and bosons.
The structure of flavor vacuum reflects into observable
quantities: exact oscillation formulas \cite{BV95} have been
derived in QFT exhibiting corrections with respect to the usual QM
ones.

In this paper, we first review some aspects of the quantization of
neutrino mixing in the case of three flavors with CP violation and then
discuss the group structure involved in the mixing and the related representations,
both for two- and three-flavor
mixing. The deformation of the associated
algebra as well as the geometric phase due to CP violation are also discussed.

\section{Three flavor fermion mixing}
\label{3flav}

We discuss here  some aspects of the QFT of three flavor  fermion mixing
\cite{BV95}.

Among the various possible parameterizations of the mixing matrix
for three fields, we choose to work with the standard
representation of the CKM matrix \cite{Cheng-Li}:
\bea\label{fermix} &&{}\hspace{-1cm} \Psi_f(x) \, = {\cal U} \,
\Psi_m (x)
\\
&&{}\hspace{-1cm} \non {\cal U}=\lf(\begin{array}{ccc}
c_{12}c_{13} & s_{12}c_{13} & s_{13}e^{-i\de} \\
-s_{12}c_{23}-c_{12}s_{23}s_{13}e^{i\de} &
c_{12}c_{23}-s_{12}s_{23}s_{13}e^{i\de} & s_{23}c_{13} \\
s_{12}s_{23}-c_{12}c_{23}s_{13}e^{i\de} &
-c_{12}s_{23}-s_{12}c_{23}s_{13}e^{i\de} & c_{23}c_{13}
\end{array}\ri),
\eea
with $c_{ij}=\cos\te_{ij}$ and  $s_{ij}=\sin\te_{ij}$, being
$\te_{ij}$ the mixing angle between $\nu_{i},\nu_{j}$ and
$\Psi_m^T=(\nu_1,\nu_2,\nu_3)$,
$\Psi_f^T=(\nu_e,\nu_\mu,\nu_\tau)$. We work here with Dirac fields although
similar conclusions are valid for Majorana neutrinos as well \cite{BV95}.

As shown in  Ref.\cite{BV95}, the generator of the
 transformation (\ref{fermix}) is\footnote{Let us
consider for example the generation of the first row of the mixing
matrix ${\cal U}$. We have  $\pa \nu_e/\pa\te_{23}\, =\, 0;$
and $ \pa\nu_e /\pa \te_{13} \, =
  \,G_{12}^{-1}G_{13}^{-1}[\nu_1,L_{13}]G_{13}G_{12}\,=\,
  G_{12}^{-1}G_{13}^{-1} e^{-i \de}\nu_3 G_{13}G_{12} $, thus:
\bea
 \non \pa^2\nu_e/\pa \te_{13}^2  \, = \, -\nu_e \quad \Rar \quad
\nu_e \, = \, f(\te_{12}) \cos\te_{13} + g(\te_{12}) \sin\te_{13};
\eea
with initial conditions: $f(\te_{12})
= \nu_e|_{\te_{13}=0}  $ and $ g(\te_{12}) = \pa\nu_e/\pa
\te_{13}|_{\te_{13}=0} = e^{-i \de}\nu_3 $.
We also have
\bea \non \pa^2f(\te_{12})/\pa \te_{13}^2  \,  = \, -f(\te_{12})
\quad \Rar \quad f(\te_{12}) \, = \, A \cos\te_{12}  \,+ B
\sin\te_{12} \eea
with the initial conditions $A=\nu_e|_{\te=0}=\nu_1$ and $B=\pa
f(\te_{12}) /\pa\te_{12}|_{\te=0}=\nu_2$, and $\te=(\te_{12},
\te_{13}, \te_{23}).$}:
\bea\label{incond} &&\nu_{\si}^{\al}(x)\equiv G^{-1}_{\bf \te}(t)
\, \nu_{i}^{\al}(x)\, G_{\bf \te}(t), \eea
with $(\si,i)=(e,1), (\mu,2), (\tau,3)$, and
\bea\label{generator} &&G_{\bf
\te}(t)=G_{23}(t)G_{13}(t)G_{12}(t)\, , \eea
where $ G_{ij}(t)\equiv \exp\Big[\te_{ij}L_{ij}(t)\Big]$ and
\bea\label{generators1} &&\hspace{-1cm} L_{12}(t)=\int d^3{\bf
x}\lf[\nu_{1}^{\dag}(x)\nu_{2}(x)-\nu_{2}^{\dag}(x)\nu_{1}(x)\ri],
\\ [2mm] \label{generators2}
&&\hspace{-1cm}L_{23}(t)=\int d^3{\bf
x}\lf[\nu_{2}^{\dag}(x)\nu_{3}(x)-\nu_{3}^{\dag}(x)\nu_{2}(x)\ri],
\\[2mm] \label{generators3}
&&\hspace{-1.4cm}L_{13}(\de,t)=\int d^3{\bf
x}\lf[\nu_{1}^{\dag}(x)\nu_{3}(x)e^{-i\de}-\nu_{3}^{\dag}(x)
\nu_{1}(x)e^{i\de}\ri]. \eea
It is clear  that the
phase $\de $ is unavoidable for three field mixing, while it can be
incorporated in the definition of the fields in the two flavor
case.

The free fields  $\nu_i$  can be quantized in the usual
way  (we use $t\equiv x_0$):
\bea\label{2.2} \nu_{i}(x) = \sum_{r} \int d^3 {\bf k} \lf[u^{r}_{{\bf
k},i}(t) \al^{r}_{{\bf k},i}\:+    v^{r}_{-{\bf k},i}(t)
\bt^{r\dag }_{-{\bf k},i}  \ri] e^{i {\bf k}\cdot{\bf x}} ,\qquad
i=1,2,3\,, \eea
with $u^{r}_{{\bf k},i}(t)=e^{-i\om_{k,i} t}u^{r}_{{\bf k},i}$,
$v^{r}_{{\bf k},i}(t)=e^{i\om_{k,i} t}v^{r}_{{\bf k},i}$ and
$\om_{k,i}=\sqrt{{\bf k}^2+m_i^2}$. The vacuum for the mass
eigenstates is denoted by $|0\ran_{m}$:  $\; \; \al^{r}_{{\bf
k},i}|0\ran_{m}= \bt^{r }_{{\bf k},i}|0\ran_{m}=0$.   The
anticommutation relations are the usual ones; the wave function
orthonormality and completeness relations are those of
Ref.\cite{BV95}.

There it was also shown that the above generator of mixing transformations
has a non-trivial action on  $|0\ran_m$.
The vacuum for the flavor fields can be then defined as:
\bea
|0(t)\ran_f \, \equiv G_{\bf \te}^{-1}(t)|0\ran_m\,.
\eea
The  flavor annihilation operators defined as  $\al^{r}_{{\bf
k},\si}(t) \equiv G^{-1}_{\bf \te}(t)\al^{r}_{{\bf k},i} G_{\bf
\te}(t)$ and $\bt^{r\dag}_{{\bf k},\si}(t)\equiv
 G^{-1}_{\bf \te}(t) \bt^{r\dag}_{{\bf k},i} G_{\bf
\te}(t)$
were studied in Ref.\cite{BV95} and shown to exhibit
a non-standard Bogoliubov like term. For example, the
annihilation operator for electron neutrino
is (in the reference frame ${\bf k}=(0,0,|{\bf
k}|)$):
\bea \non
\al_{{\bf k},e}^{r}(t)&=&c_{12}c_{13}\;\al_{{\bf k},1}^{r} +
s_{12}c_{13}\lf(U^{{\bf k}*}_{12}(t)\;\al_{{\bf k},2}^{r}
+\epsilon^{r} V^{{\bf k}}_{12}(t)\;\bt_{-{\bf k},2}^{r\dag}\ri)
\\
&& + e^{-i\de}\;s_{13}\lf(U^{{\bf k}*}_{13}(t)\;\al_{{\bf k},3}^{r}
+\,\epsilon^{r} V^{{\bf k}}_{13}(t)\;\bt_{-{\bf k},3}^{r\dag}\ri)\;,
\eea
with Bogoliubov coefficients defined as:
\bea
&&V^{{\bf k}}_{ij}(t)=|V^{{\bf
k}}_{ij}|\;e^{i(\om_{k,j}+\om_{k,i})t}\;\;\;\;,\;\;\;\; U^{{\bf
k}}_{ij}(t)=|U^{{\bf k}}_{ij}|\;e^{i(\om_{k,j}-\om_{k,i})t}
\\
&&|U^{{\bf
k}}_{ij}|=\lf(\frac{\om_{k,i}+m_{i}}{2\om_{k,i}}\ri)
^{\frac{1}{2}}
\lf(\frac{\om_{k,j}+m_{j}}{2\om_{k,j}}\ri)^{\frac{1}{2}}
\lf(1+\frac{|{\bf k}|^{2}}{(\om_{k,i}+m_{i})
(\om_{k,j}+m_{j})}\ri)
\\
&&|V^{{\bf k}}_{ij}|=\lf(\frac{\om_{k,i}+m_{i}}{2\om_{k,i}}\ri)
^{\frac{1}{2}}
\lf(\frac{\om_{k,j}+m_{j}}{2\om_{k,j}}\ri)^{\frac{1}{2}}
\lf(\frac{|{\bf k}|}{(\om_{k,j}+m_{j})}-\frac{|{\bf
k}|}{(\om_{k,i}+m_{i})}\ri)
\eea
where $i,j=1,2,3$ and $j>i$. We also have
$|U^{{\bf k}}_{ij}|^{2}+|V^{{\bf k}}_{ij}|^{2}=1$.

The flavor fields can be expanded in terms of the flavor ladder operators as:
\bea\label{exnue1} &&\nu_\si(x)= \sum_{r} \int d^3 {\bf k}
\lf[ u^{r}_{{\bf k},i}(t) \al^{r}_{{\bf k},\si}(t) +
v^{r}_{-{\bf k},i}(t) \bt^{r\dag}_{-{\bf k},\si}(t) \ri]  e^{i
{\bf k}\cdot{\bf x}}, \eea
with $(\si,i)=(e,1), (\mu,2), (\tau,3)$.

Let us now investigate the
algebraic structures associated with the mixing generator Eq.(\ref{generator}).
To this end we introduce the following Lagrangian:
\bea
\label{lag12}
{\cal L}(x)&=&  {\bar \Psi_m}(x) \lf( i
\not\!\partial -  \textsf{M}_d\ri) \Psi_m(x)
\;=\; {\bar \Psi_f}(x) \lf( i
\not\!\partial - \textsf{M} \ri) \Psi_f(x)\, ,
\eea
where $\textsf{M}_d = diag(m_1,m_2,m_3)$ and the matrix $ \textsf{M}$ is non-diagonal,
being fixed by the mixing relations Eq.(\ref{fermix}).

The above Lagrangian is invariant under global $U(1)$ phase transformations,
leading to a conserved (total) charge $Q=
\int  d^3{\bf x}\,\Psi^\dag_m(x)\,\Psi_m(x)\,=\,
\int  d^3{\bf x}\,\Psi^\dag_f(x)\,\Psi_f(x)$.

We then study the invariance of ${\cal L}$ under global
phase transformations of the kind:
\bea \label{masssu3} \Psi_m'(x) \, =\, e^{i \al_j  {\ti F}_j}\, \Psi_m
(x) \, \qquad, \qquad
 j=1, 2,..., 8.
\eea
where ${\ti F}_{j}\equiv\frac{1}{2}{\ti \lambda}_{j}$ and the
${\ti \lambda}_{j}$ are a generalization of the usual Gell-Mann
matrices $\lambda_{j}$:
\bea \non &&{}\hspace{-.5cm} {\ti \lambda}_{1}=\lf(\ba{ccc}
  0 & e^{i\de_2} & 0 \\
 e^{-i\de_2} & 0 & 0 \\
  0 & 0 & 0
\ea\ri) \;,\quad {\ti \lambda}_{2}=\lf(\ba{ccc}
  0 & -i e^{i\de_2} & 0 \\
  i e^{-i\de_2}& 0 & 0 \\
  0 & 0 & 0
\ea \ri)\;,\quad
{\ti \lambda}_{3}=\lf(\ba{ccc}
  1 & 0 & 0 \\
  0 & -1 & 0 \\
  0 & 0 & 0
\ea\ri)\;
\\[2mm] \label{gelm}
&&{}\hspace{-.5cm} {\ti \lambda}_{4}=\lf(\ba{ccc}
  0 & 0 & e^{-i\de_5}\\
  0 & 0 & 0 \\
  e^{i\de_5} & 0 & 0
\ea\ri)\;,\qquad {\ti \lambda}_{5}=\lf(\ba{ccc}
  0 & 0 & -ie^{-i\de_5} \\
  0 & 0 & 0 \\
  ie^{i\de_5} & 0 & 0
\ea\ri)
\\ [2mm] \non
&&{}\hspace{-.5cm} {\ti \lambda}_{6}=\lf(\ba{ccc}
  0 & 0 & 0 \\
  0 & 0 & e^{i\de_7} \\
  0 & e^{-i\de_7} & 0
\ea\ri)\;, \quad {\ti \lambda}_{7}=\lf(\ba{ccc}
  0 & 0 & 0 \\
  0 & 0 & -i e^{i\de_7}\\
  0 & ie^{-i\de_7} & 0
\ea\ri)
 \;,\quad {\ti \lambda_{8}}=\frac{1}{\sqrt{3}}\lf(\ba{ccc}
  1 & 0 & 0 \\
  0 & 1 & 0 \\
  0 & 0 & -2
\ea\ri). \eea
The normalization is  $tr(  {\ti \lambda}_{j} {\ti
\lambda}_{k}) =2\delta_{jk}$.
One then obtains the following set of charges
\cite{BV95}:
\bea\label{su3charges} &&{}\hspace{-.5cm} {\ti Q}_{m,j}(t)\, =\,
\int  d^3{\bf x}\,\Psi^\dag_m(x)\, {\ti F}_j\, \Psi_m(x) \;,
\qquad j=1, 2,..., 8.
\eea
Thus the matrix Eq.(\ref{fermix}) is
generated by ${\ti Q}_{m,2}(t)$, ${\ti Q}_{m,5}(t)$ and ${\ti
Q}_{m,7}(t)$, with $\{\de_2,\de_5,\de_7\}\rar \{0,\de,0\}$.
An interesting point is that the algebra generated by the
matrices Eq.(\ref{gelm}) {\em is not} $su(3)$ unless the condition
$\De \equiv \de_2+\de_5 +\de_7 =0$ is imposed: such a condition is
however incompatible with the presence of a CP violating phase.
When CP violation is allowed, then $\De \neq 0$ and the $su(3)$
algebra is deformed. To see this, let us introduce the raising and
lowering operators, defined as \cite{Cheng-Li}:
\bea && {\ti T}_\pm \equiv {\ti F}_1 \pm i {\ti F}_2 \quad, \quad
{\ti U}_\pm \equiv {\ti F}_6 \pm i {\ti F}_7 \quad, \quad {\ti
V}_\pm \equiv {\ti F}_4 \pm i {\ti F}_5 \eea
We also define:
\bea && {}\hspace{-.5cm}
{\ti Y}= \frac{2}{\sqrt{3}}{\ti F}_8 \quad, \quad {\ti
T}_3 \equiv {\ti F}_3 \quad, \quad {\ti U}_3 \equiv
\frac{1}{2}\lf(\sqrt{3} {\ti F}_8 - {\ti F}_3 \ri) \quad, \quad
{\ti V}_3 \equiv \frac{1}{2}\lf(\sqrt{3} {\ti F}_8 + {\ti F}_3
\ri) \eea
The commutation relations are
\bea \lab{co1} && {}\hspace{-.5cm}
[{\ti T}_3,{\ti T}_\pm] \,=\, \pm {\ti T}_\pm \;, \quad  [{\ti
T}_3,{\ti U}_\pm] \,=\, \mp \frac{1}{2} {\ti U}_\pm \;, \quad
[{\ti T}_3,{\ti V}_\pm] \,=\, \pm \frac{1}{2} {\ti V}_\pm \;, \quad
[{\ti T}_3,{\ti Y}] \,=\, 0 \;, \quad
\\ [2mm]
&&  {}\hspace{-.5cm}
[{\ti Y},{\ti T}_\pm] \,=\, 0 \;, \quad [{\ti Y},{\ti
U}_\pm] \,=\, \pm  {\ti U}_\pm \;, \quad [{\ti Y},{\ti V}_\pm]
\,=\, \pm  {\ti V}_\pm \;,\quad
[{\ti T}_+,{\ti T}_-] \,=\, 2 {\ti T}_3 \;, \quad
\\[2mm]
&&  {}\hspace{-.5cm}
[{\ti U}_+,{\ti U}_-] \,=\, 2 {\ti U}_3 \;, \quad [{\ti V}_+,{\ti
V}_-] \,=\, 2 {\ti V}_3 \;, \quad[{\ti T}_+,{\ti V}_+]
\,=\, [{\ti T}_+,{\ti U}_-] \,=\,[{\ti
U}_+,{\ti V}_+] \,=\, 0,
 \eea
that are similar to the standard $SU(3)$ commutation relations. However,
the following commutators are deformed:
\bea \lab{co2} &&{}\hspace{-.5cm} [{\ti T}_+,{\ti V}_-] \,=\, -
{\ti U}_- \,e^{2 i \De {\ti U}_3} \quad, \quad [{\ti T}_+,{\ti
U}_+] \,=\, {\ti V}_+ \,e^{-2 i \De {\ti V}_3} \quad, \quad [{\ti
U}_+,{\ti V}_-] \,=\, {\ti T}_- \,e^{2 i \De {\ti T}_3}
\eea
%
Let us define the operators:
\bea\lab{noether1} Q_{1}& \equiv &\frac{1}{3}Q \,+ \,Q_{m,3}+
\,\frac{1}{\sqrt{3}}Q_{m,8},
\\ \lab{noether2}
Q_{2}& \equiv & \frac{1}{3}Q \,-
\,Q_{m,3}+\,\frac{1}{\sqrt{3}}Q_{m,8},
\\ \lab{noether3}
Q_{3}& \equiv &\frac{1}{3}Q \,- \,\frac{2}{\sqrt{3}}Q_{m,8}, \eea
\bea\lab{charge}
 &&Q_i \, = \,\sum_{r} \int d^3 k\lf(
\al^{r\dag}_{{\bf k},i} \al^{r}_{{\bf k},i}\, -\,
\bt^{r\dag}_{-{\bf k},i}\bt^{r}_{-{\bf k},i}\ri),\,\quad  i=1, 2, 3 .
\eea
These are nothing but  the Noether charges associated with the
non-interacting fields $\nu_1$, $\nu_2$ and $\nu_3$: in the
absence of mixing, they are the flavor charges,  separately
conserved for each generation.

In a similar way with the above derivation, we can study the invariance
properties of the Lagrangian Eq.(\ref{lag12}) under the
transformations:
\bea \label{flavsu3} \Psi_f'(x) \, =\, e^{i \al_j  {\ti F}_j}\, \Psi_f
(x) \, , \qquad  j=1, 2,..., 8.
\eea
Then the following charges are obtained
\bea\label{su3fcharges} &&{}\hspace{-.5cm} {\ti Q}_{f,j}(t)\, =\,
\int  d^3{\bf x}\,\Psi^\dag_f(x)\, {\ti F}_j\, \Psi_f(x) \;,
\qquad j=1, 2,..., 8.
\eea
In contrast with the previous case, note that  the diagonal
elements ${\ti Q}_{f,3}$ and
${\ti Q}_{f,8}$ are now time-dependent.
We define the {\em flavor
charges} for mixed fields as
\bea Q_e(t) & \equiv & \frac{1}{3}Q \, + \, Q_{f,3}(t)\, +
\,\frac{1}{\sqrt{3}} Q_{f,8}(t),
\\
Q_\mu(t) & \equiv & \frac{1}{3}Q \, - \, Q_{f,3}(t)+
\,\frac{1}{\sqrt{3}} Q_{f,8}(t),
\\
Q_\tau(t) & \equiv & \frac{1}{3}Q \, -  \, \frac{2}{\sqrt{3}}
Q_{f,8}(t). \eea
with $Q_e(t) \, + \,Q_\mu(t) \,+ \,Q_\tau(t) \, = \, Q$.
These charges have a simple expression in terms of the flavor
ladder operators:
\bea\lab{flavchar} Q_\si(t) & = & \sum_{r} \int d^3 k\lf(
\al^{r\dag}_{{\bf k},\si}(t) \al^{r}_{{\bf k},\si}(t)\, -\,
\bt^{r\dag}_{-{\bf k},\si}(t)\bt^{r}_{-{\bf
k},\si}(t)\ri),\quad \si= e,\mu,\tau, \eea
because of the connection with the Noether charges of
Eq.(\ref{charge}) via the mixing generator: $Q_\si(t) =
G^{-1}_\te(t)Q_i G_\te(t)$.
In Ref.\cite{3flavors} the above flavor charges were used to derive
oscillation formulas which generalize the usual ones obtained in Quantum
Mechanics.

\section{Group representations and the oscillation formula}

We now study the group representations. Let us first consider
the simple case of two generations and then discuss the three flavor case.

\subsection{Two flavors}

In this case \cite{BV95}, the group is $SU(2)$
and the charges in the mass basis read \cite{BV95}:
\bea
Q_{m,j}(t)\,=\,\frac{1}{2}\int d^{3}{\bf x}\,   \Psi_m^\dag(x) \, \tau_j\,
\Psi_m(x),\quad \qquad  j \,=\, 1, 2, 3,
\eea
where $\Psi_m^T=(\nu_1,\nu_2)$ and  $\tau_j=\si_j/2$ with $\si_j$ being
the Pauli matrices.

The states with definite masses can then be defined as eigenstates
of $Q_{m,3}$:
\bea \label{eigsu2a}
Q_{m,3}|\nu_{1}\rangle=\frac{1}{2}|\nu_{1}\rangle\qquad ;\qquad
Q_{m,3}|\nu_{2}\rangle=-\frac{1}{2}|\nu_{2}\rangle
\eea
and similar ones for antiparticles. We have $|\nu_i\ran =
\al^{r\dag}_{{\bf k},i} |0\ran_m$, $i=1,2$. Eq.(\ref{eigsu2a}) expresses the obvious
fact that the mass eigenstates, treated as free particle states, are eigenstates of the
conserved $U(1)$ charges associated to $\nu_1$ and $\nu_2$:
\bea\label{su2noether}
Q_{1}\, \equiv \,\frac{1}{2}Q \,+ \,Q_{m,3}
\qquad; \qquad Q_{2}\, \equiv \,\frac{1}{2}Q \,- \,Q_{m,3}.
\eea

The next step is to define flavor states using a similar procedure. We need to be
careful here since the diagonal $SU(2)$ generator $Q_{f,3}$ is time-dependent in the
flavor basis. Thus we define states (Hilbert space) at a reference time $t=0$ from:
\bea
Q_{f,3}(0)|\nu_{e}\rangle=\frac{1}{2}|\nu_{e}\rangle\quad;\quad
Q_{f,3}(0)|\nu_{\mu}\rangle=-\frac{1}{2}|\nu_{\mu}\rangle.
\eea
with $|\nu_\si\ran = \al^{r\dag}_{{\bf k},\si}(0) |0(0)\ran_f$,
$\si = e,\mu$ and similar ones for antiparticles.

The flavor states so defined are  eigenstates of the flavor
charges at time $t=0$:
\bea
&&Q_{e}(t)=\frac{1}{2}Q + Q_{f,3}(t)\qquad; \qquad
Q_{\mu}(t)=\frac{1}{2}Q - Q_{f,3}(t),
\\ [2mm] \label{su2flavstates}
&& Q_{e}(0)|\nu_{e}\rangle=|\nu_{e}\rangle\qquad ;\qquad
Q_{\mu}(0)|\nu_{\mu}\rangle=|\nu_{\mu}\rangle.
\eea
and $Q_{e}(0)|\nu_\mu\rangle=Q_{\mu}(0)|\nu_e\rangle=0$.

This result is far from being trivial since
the usual Pontecorvo states \cite{Pont}:
\begin{eqnarray} \label{nue0a}
|\nu_{e}\rangle_P &=& \cos\theta\;|\nu_{1}\rangle \;+\;
\sin\theta\; |\nu_{2}\rangle \;
\\ [2mm] \label{nue0b}
|\nu_{\mu}\rangle_P &=& -\sin\theta\;|\nu_{1}\rangle \;+\;
\cos\theta\; |\nu_{2}\rangle \; ,
\end{eqnarray}
are {\em not} eigenstates of the flavor charges. The Lorentz invariance properties of
the flavor states Eq.(\ref{su2flavstates}) have been discussed in Ref.\cite{dispersion}.

At a  time $t\neq 0$, oscillation formulas can be derived \cite{BV95} for the
flavor charges from the following relation
\be \langle\nu_{e}|Q_{f,3}(t)|\nu_{e}\rangle = \frac{1}{2} -
|U^{\bf k}_{12}|^{2}\sin^{2}(2\theta)
\sin^{2}\lf(\frac{\omega_{k,2}-\omega_{k,1}}{2}t\ri)
-|V^{\bf k}_{12}|^{2}\sin^{2}(2\theta)
\sin^{2}\lf(\frac{\omega_{k,2}+\omega_{k,1}}{2}t\ri)
\ee
where the non-standard oscillation term do appear.

\subsection{Three flavors}

\begin{figure}[t]
\setlength{\unitlength}{1mm} \vspace*{70mm} 
\includegraphics{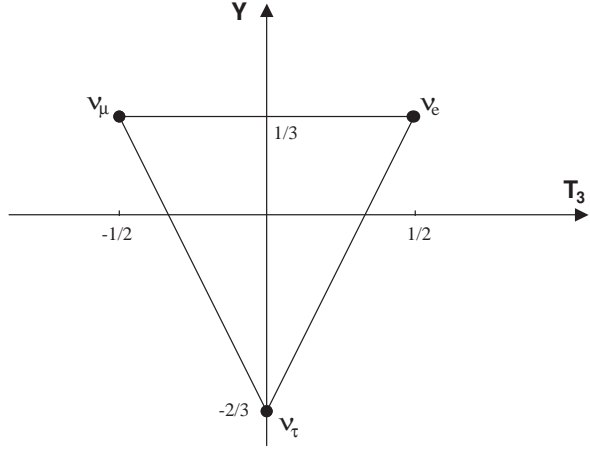}
\caption{The triplet.}

\mlab{fig4}
\end{figure}

Having discussed above the  procedure for the definition of flavor states in
the case of two flavors,
we can directly write, for three flavors,
\bea
Q_{\si}(0)|\nu_{\si}\rangle=|\nu_{\si}\rangle \quad,
\qquad Q_{\si}(0)|{\bar \nu}_{\si}\rangle=-|{\bar \nu}_{\si}\rangle
\quad , \qquad \si=e,\mu,\tau,
\eea
leading to
\bea
|\nu_\si\ran \equiv \al^{r\dag}_{{\bf k},\si}(0) |0(0)\ran_f\quad,
\qquad |{\bar \nu}_\si\ran \equiv \bt^{r\dag}_{{\bf k},\si}(0) |0(0)\ran_f\quad,
\qquad \si=e,\mu,\tau.
\eea

These neutrino and antineutrino
states can be related to the fundamental representation  ${\bf 3}$ and
${\bf 3^*}$ of the
(deformed) $SU(3)$ mixing group above introduced, as shown in Fig.1 for
neutrinos.
Note that the position of the points in the ${\ti Y}-{\ti T}_{3}$ is the same
as for the ordinary $SU(3)$, since the diagonal matrices ${\ti \lambda_{3}}$,
${\ti \lambda_{8}}$ do not contain phases.
However, a closed loop around the triangle gives a non-zero phase which is of
geometrical origin \cite{Anandan}
and only depends on the CP phase. A similar situation is
valid for antineutrinos.

\vspace{0.5cm}

To see this more in detail, let us consider the octet representation
as in Fig.2 and define the normalized state $|A\rangle$: $ \langle A
|A\rangle =1$. Then all the other states are also normalized,
except for  $|G\rangle$: $|G\rangle = \frac{1}{\sqrt{2}}\ti{T}_{-}|A\rangle$.
%
\begin{figure}[t]
\setlength{\unitlength}{1mm} \vspace*{70mm} 
\includegraphics{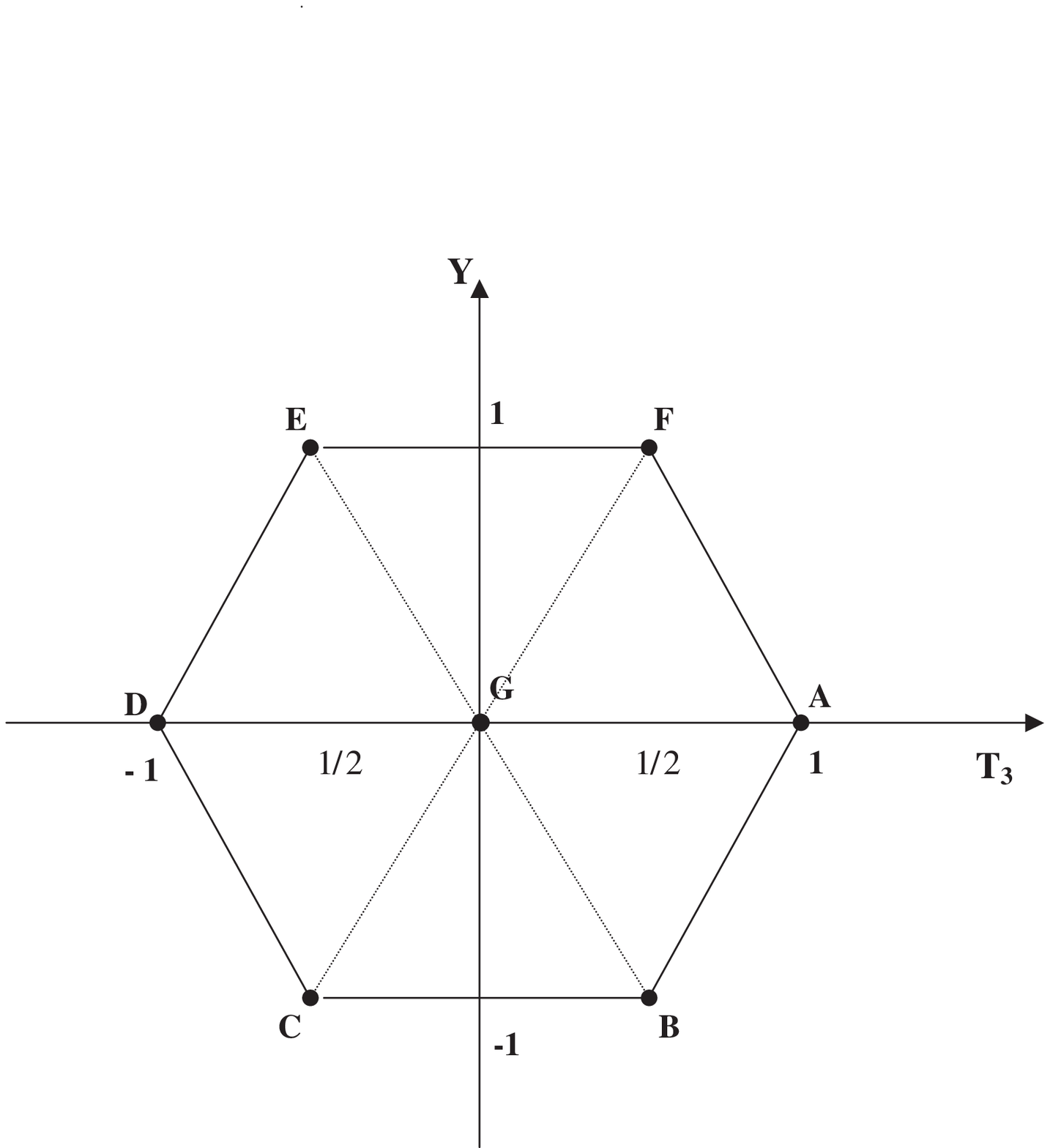}
\caption{The octet.}

\mlab{fig3}
\end{figure}
We obtain the following paths
\bea\non
 &(AGBA): &\ti{V}_{+}\ti{U}_{-}\ti{T}_{-}|A\rangle =
\ti{V}_{+}([\ti{U}_{-},\ti{T}_{-}] +
\ti{T}_{-}\ti{U}_{-})|A\rangle =\ti V_{+}\ti V_{-}e^{2i\Delta \ti
V_{3}}|A\rangle = e^{i\Delta}|A\rangle
\\[2mm] \non
&(ABGA): &\ti T_{+}\ti U_{+}\ti V_{-}|A\rangle = e^{-i\Delta}|A\rangle
\\[2mm] \non
&(AFGA): &\ti T_{+}\ti V_{-}\ti U_{+}|A\rangle
= -e^{-i\Delta}|A\rangle
\\[2mm] \non
&(AGFA): &\ti
U_{-}\ti V_{+}\ti T_{-}|A\rangle = -e^{i\Delta}|A\rangle
\\[2mm] \non
&(AFGBA): &\ti V_{+}\ti U_{-}\ti V_{-}\ti U_{+}|A\rangle
= |A\rangle
\\[2mm]
&(AFEDCBA): &\ti V_{+}\ti T_{+} \ti
U_{-}\ti V_{-}\ti T_{-}\ti U_{+}|A\rangle = |A\rangle
\eea
where
we have used
\bea
&&{}\hspace{-1.5cm}
\ti U_{-}|A\rangle = \ti T_{+}|A\rangle = \ti
V_{+}|A\rangle = 0,
\quad\ti T_{3}|A\rangle = |A\rangle \;,
\quad \ti V_{3}|A\rangle = \frac{1}{2}|A\rangle \;, \quad \ti
U_{3}|A\rangle = - \frac{1}{2}|A\rangle
\eea
and the commutation relations.
%
%

We thus see  that the phase sign change if we change the versus
of the path on the triangles; the paths on two opposite triangles
and around the hexagon bring no phase.

\section{Conclusions}

In this report, we have discussed some aspects of
the quantization of
mixed fermions (neutrinos) in the context of Quantum Field Theory.

In particular, we have analyzed the algebraic structures arising in
connection with field mixing and their deformation due to the presence
of CP violating phase, in the case of neutrino mixing among three generations.

We have defined flavor states in terms of the  representations of the group associated
with field mixing.
A new geometric phase arising from  CP violation was discovered. Other geometric
phases related to fermion mixing have been discussed in Refs.\cite{berry}.

\section*{Acknowledgements}
M.B. and G.V. are grateful to the organizers of the
XII-th International Baksan School ``Particles
and Cosmology" for the invitation.
The ESF Program COSLAB, EPSRC, INFN and
INFM are also acknowledged for partial financial support.

\end{document}